\documentclass[a4paper]{spie}  %>>> use this instead for A4 paper
\usepackage{amsmath,amsfonts,amssymb}
\usepackage{graphicx}
\usepackage[colorlinks=true, allcolors=blue]{hyperref}

\title{The MICADO first light imager for the ELT: \\ off-axis performance of PSF reconstruction}

\author[a]{Matteo Simioni}
\author[b]{Daniel Jodlbauer}
\author[a]{Carmelo Arcidiacono}
\author[a]{Andrea Grazian}
\author[a]{Marco Gullieuszik}
\author[c]{Elisa Portaluri}
\author[a]{Benedetta Vulcani}
\author[d]{Roland Wagner}
\author[a]{Anita Zanella}
\author[e,f]{Johanna Hartke}
\author[g]{Tapio Helin}
\author[f]{Hanindyo Kuncarayakti}
\author[h]{Elena Masciadri}
\author[i]{Fernando Pedichini}
\author[i]{Roberto Piazzesi}
\author[h]{Alessio Turchi}
\author[i]{Piero Vaccari}
\affil[a]{INAF – Osservatorio Astronomico di Padova, Vicolo Osservatorio 5, I-35122, Padova, Italy}
\affil[b]{Industrial Mathematics Institute, Johannes Kepler University, Altenberger Stra{\ss}e 69, Linz, 4040, Linz, Austria}
\affil[c]{INAF – Osservatorio Astronomico d'Abruzzo, Via Mentore Maggini, I-64100, Teramo, Italy}
\affil[d]{Johann Radon Institute for Computational and Applied Mathematics of the Austrian Academy of Sciences (RICAM), Altenberger Stra{\ss}e 69, 4040, Linz, Austria}
\affil[e]{Finnish Centre for Astronomy with ESO (FINCA), University of Turku, Finland}
\affil[f]{University of Turku, Turun yliopisto, FI-20014, Turku, Finland}
\affil[g]{LUT University, Yliopistonkatu 34, 53850, Lappeenranta, Finland}
\affil[h]{INAF-Osservatorio Astrofisico di Arcetri, L.go Enrico Fermi 5, I - 50125, Firenze, Italy}
\affil[i]{INAF - Osservatorio astronomico di Roma, Via Frascati 33, I-00078, Monte Porzio Catone, Italy}

\authorinfo{Further author information: (Send correspondence to M.S.)\\M.S.: E-mail: matteo.simioni@inaf.it}

\begin{document} 
\maketitle

\begin{abstract}
The highest scientific return, for adaptive optics (AO) observations, is achieved with a reliable reconstruction of the PSF. This is especially true for MICADO@ELT. 
In this presentation, we will focus on extending the MICADO PSF reconstruction  (PSF-R) method to the off-axis case. Specifically, a novel approach based on temporal-based tomography of AO telemetry data has been recently implemented. Results from the PSF-R of both simulated and real data show that, at half isoplanatic angle distances, a precision of about $10-15\%$ is achievable in both Strehl ratio and full-width at half maximum, paving the way to extend the MICADO PSF-R tool also to the multi-conjugated AO case.

\end{abstract}

\keywords{Adaptive optics, Point spread functions, Point spread function reconstruction, Infrared imaging, Astronomy, MICADO@ELT, MICADO PSF reconstruction, ERIS@VLT PSF reconstruction}

\section{INTRODUCTION}\label{sec:intro}  % \label{} allows reference to this section

MICADO\cite{2016SPIE.9908E..1ZD} will be the first-light adaptive optics (AO) assisted imager of the ESO extremely large telescope. When integrated, it will operate in single-conjugated AO (SCAO) mode. Multi-conjugate AO  (MCAO) observations will be subsequently possible with MICADO when MORFEO\cite{2022SPIE12185E..14C} will start operations. It will also be the first ESO instrument designed to offer a PSF reconstruction (PSF-R) service to the final user: the detailed knowledge of the PSF is crucial for the majority of the planned scientific drivers\cite{2020SPIE11448E..37S,2022SPIE12185E..41G}. 
In this context, the core algorithm of the MICADO PSF-R tool\cite{2018JATIS...4d9003W} follows a telemetry-only approach and the PSF is reconstructed without using information from the science images. It will rely instead on the analysis of the AO telemetry, recorded simultaneously with the scientific observations, on careful calibrations, and on additional external information (e.g. regarding the turbulence profile).

The MICADO PSF-R tool was proven to work once applied to real SCAO on-axis observations (e.g. reconstructing SOUL+LUCI@LBT data\cite{2022JATIS...8c8003S}), consistently achieving precision in the reconstruction of the on-axis PSF of the order of $2-5\%$ in both Strehl Ratio (SR) and full-width at half maximum (FWHM) and highlighting the versatility of the method. 

However, the highest scientific return, especially for SCAO observations, is achieved with a reliable reconstruction of the off-axis PSF. This capability has been recently implemented in the MICADO PSF-R algorithm enabling the derivation of the residual incoming wavefront in any selected off-axis direction through a temporal-based tomography of AO telemetry data\cite{2023JOSAA..40.1382W}. 

Our off-axis method was evaluated using both simulated\cite{2023JOSAA..40.1382W,2023aoel.confE..34S} and real SCAO data. This presentation will focus on validating our PSF-R method using real data, specifically SCAO observations of off-axis point sources. 
In addition, a scientific evaluation has been performed, using the simulated off-axis PSFs to quantify the impact of our PSF-R, on a selected, idealized scientific case: the morphology of compact star-forming regions in high-redshift galaxies.

\section{DATA}\label{sec:data}
The data used consists of ERIS@VLT\cite{2023A&A...674A.207D} SCAO observations of a bright asterism, obtained in the context of a dedicated ESO technical time request (TTR). Three bright stars are present in the observed field of view (FoV), one of them has been used as the AO reference star (i.e. has been observed on-axis) while the other $2$ have a radial distance of $7"$ and $16"$ respectively. 
%FIG 1
\begin{figure} [tb]
   \begin{center}
   \begin{tabular}{c} %% tabular useful for creating an array of images 
   \includegraphics[height=12.5cm]{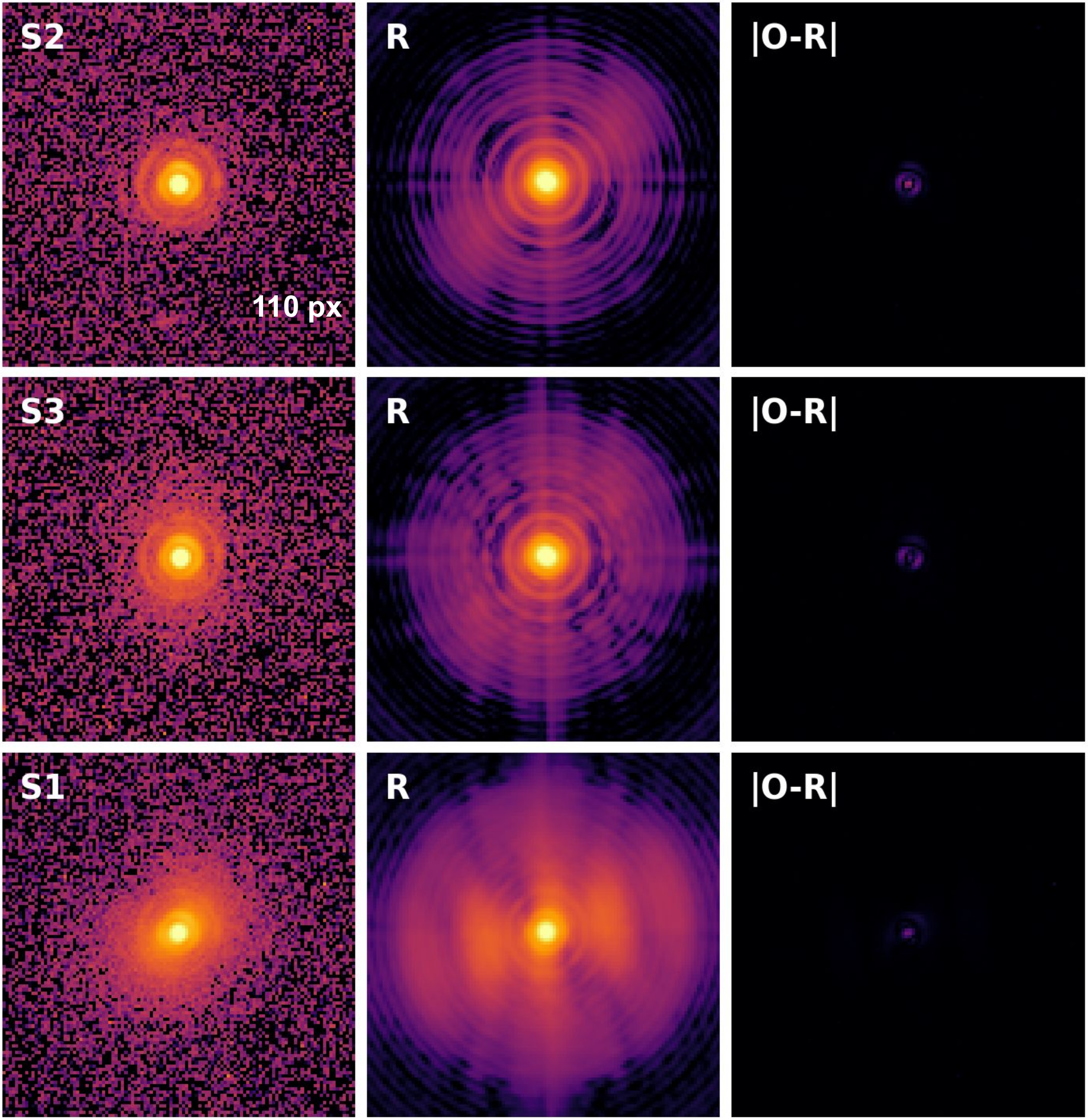}
   \end{tabular}
   \end{center}
   \caption[PSFs]{ \label{fig:psfs} Comparison between observed (left column) and reconstructed (middle column) ERIS@VLT PSFs; Rows are associated with increasing off-axis distance from top to bottom: S2 is the AO reference star, S3 is the $7"$ off-axis one while S1 has an off-axis distance of $16"$. PSFs are shown in logarithmic color scale, residuals ($|$O-R$|$, right column) in linear color scale. The box width is $110$px ($1".4$). }
\end{figure} 
%%%%%%
The observed PSFs elongation is proportional to off-axis distance and is mostly radially aligned with the AO reference star. The different elongation of the observed PSFs can be appreciated in the left panels of Figure~\ref{fig:psfs}.
Magnitudes are in the range H$=9-9.5$ VEGAmag and the observations have been performed in Fe-II narrow-band filter (${\rm \lambda_c}=1.644{\rm \mu m}$). We analyzed $225$s of continuous observations subdivided in frames of $1$s each. The data is accompanied by the associated AO telemetry, saved simultaneously to scientific observations\cite{2018JATIS...4d9003W,2022JATIS...8c8003S,2023JOSAA..40.1382W} with a frequency of $1$KHz. 

In addition to ERIS data, a sample of simulated MICADO@ELT SCAO off-axis PSFs has been considered, the associated reconstructed PSFs have already been presented in Ref.~\citenum{2023aoel.confE..34S}. They are $3$ simulated 30" off-axis SCAO PSFs at $2.2{\rm \mu m}$ (e.g. K band) with different AO reference brightness regimes.  

\section{DISCUSSION}\label{sec:discussion}

The tomographic approach of the MICADO PSF-R\cite{2023JOSAA..40.1382W} requires an estimation of the atmospheric turbulence and the wind vector profiles, in addition to the information already needed for the on-axis reconstruction\cite{2018JATIS...4d9003W,2022JATIS...8c8003S}. In the present reconstruction, we integrated MASS-DIMM measurements with accurate wind model predictions\cite{2013MNRAS.436.1968M}, to obtain a sampling of the atmosphere of $3$ layers at the height of $0.1$, $1$, and $10$ km. The resulting reconstructed PSFs are reported in Figure~\ref{fig:psfs} (middle column). 
%FIG 2
\begin{figure} [tb]
   \begin{center}
   \begin{tabular}{c} %% tabular useful for creating an array of images 
   \includegraphics[width=.3\textwidth,trim={4.5cm 0 4cm 0},clip]{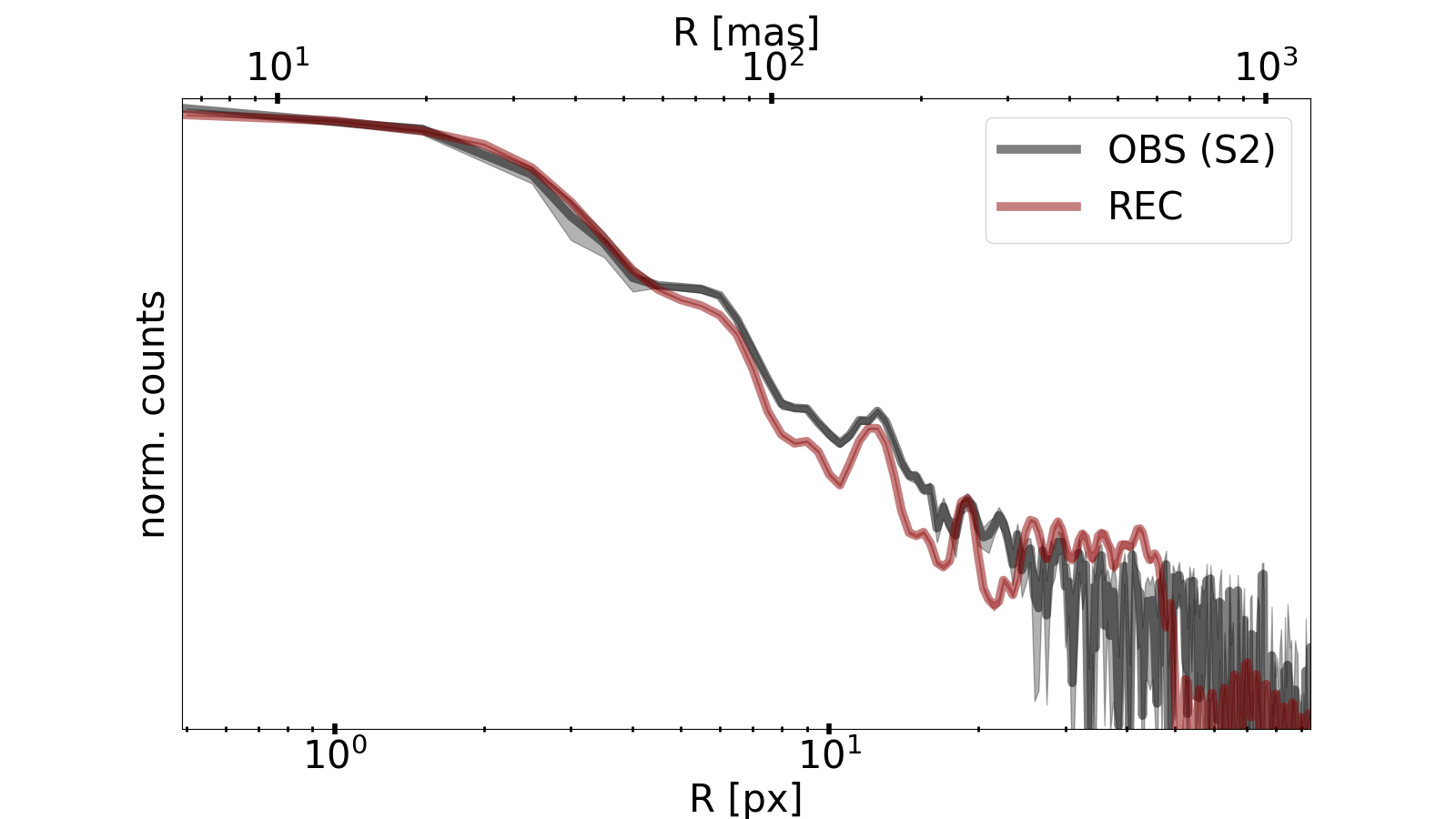}
   \includegraphics[width=.3\textwidth,trim={4.5cm 0 4cm 0},clip]{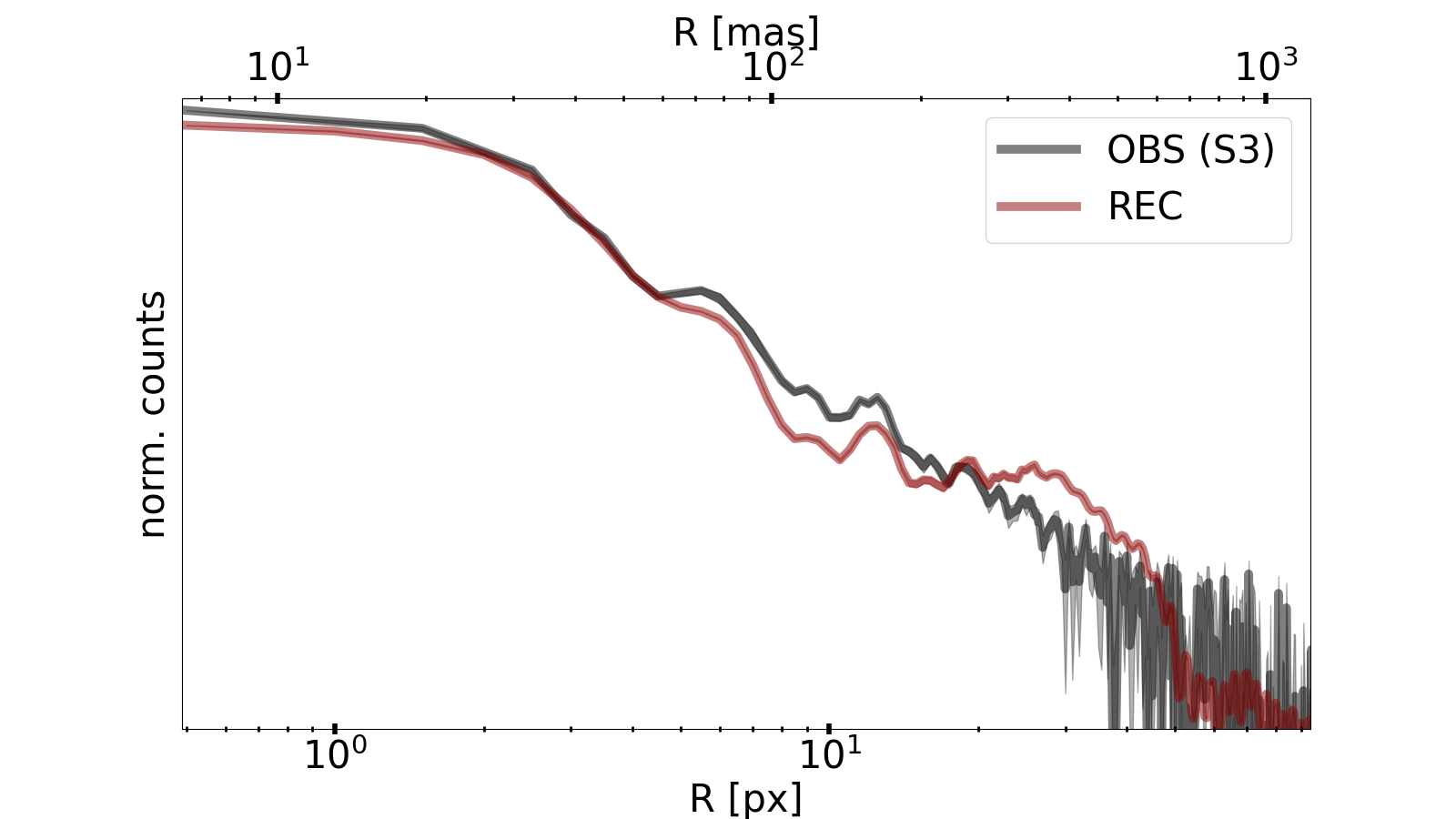}
   \includegraphics[width=.3\textwidth,trim={4.5cm 0 4cm 0},clip]{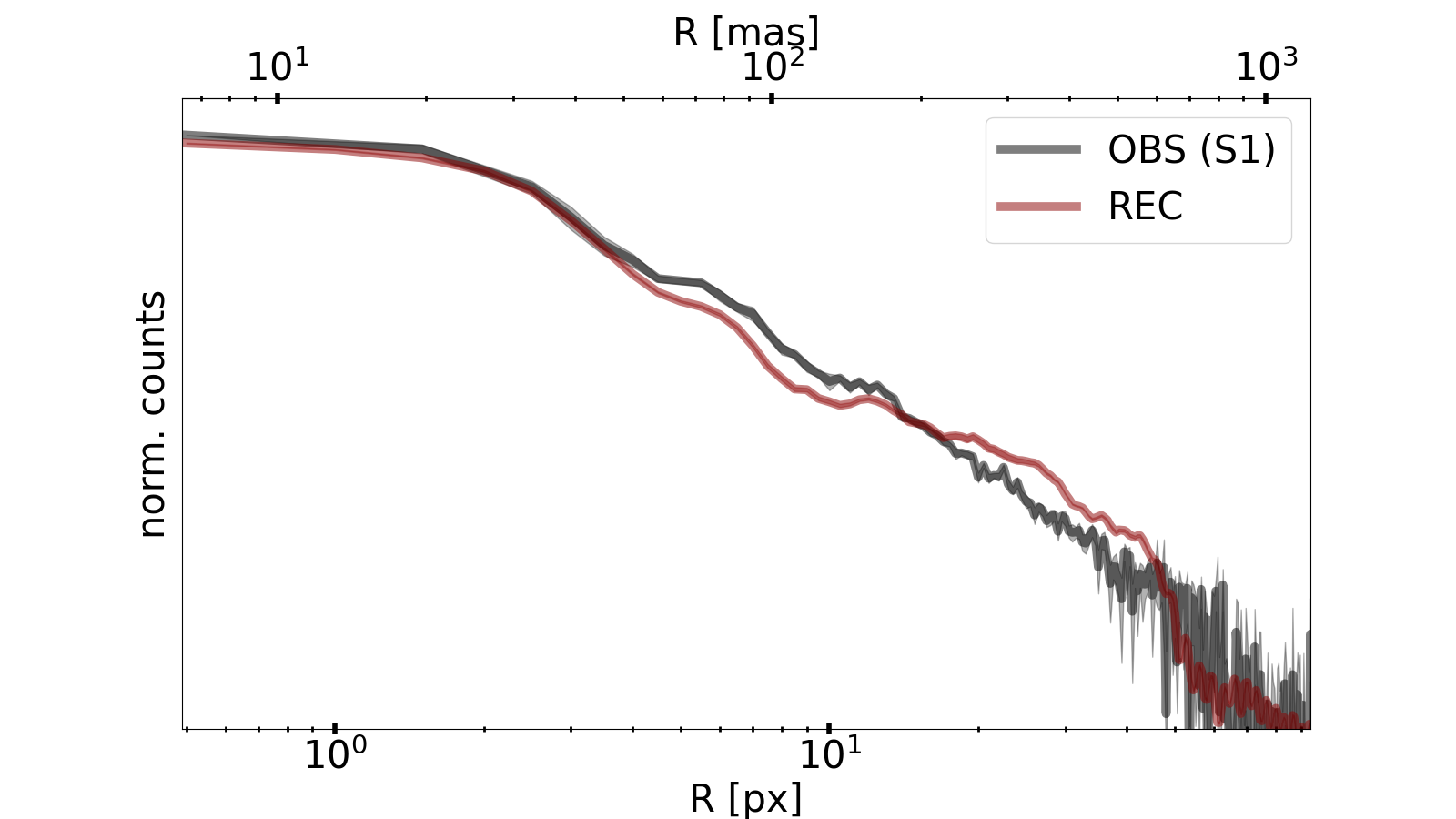}
   \end{tabular}
   \end{center}
   \caption[Radial profiles]{ \label{fig:radp} Comparison between the radial profiles of observed PSFs (black lines) and of the reconstructed ones (red lines). Off-axis distance is increasing from left to right, the name convention for the PSFs is the same as in Figure~\ref{fig:psfs}. A logarithmic scale has been used for the luminosity and the radial distance. All the profiles are normalized in flux on a circular aperture of radius $100$px ($1".3$, maximum radial distance shown). As a reference, a signal-to-noise ratio of $5$ for the observed PSFs is also marked.}
\end{figure} 
%%%%%%
From the residuals, rightmost column of Figure~\ref{fig:psfs}, the good agreement between each pair of observed and reconstructed PSFs can be noted. This can be further quantified from Figure~\ref{fig:radp} where the radial plots are shown with increasing off-axis distance from left to right. All the profiles are normalized in flux and the same scale has been used for the $3$ stars highlighting the decrease of SR at increasing off-axis distance.

\begin{figure} [tb]
   \begin{center}
   \begin{tabular}{c} %% tabular useful for creating an array of images 
   \includegraphics[width=.7\textwidth]{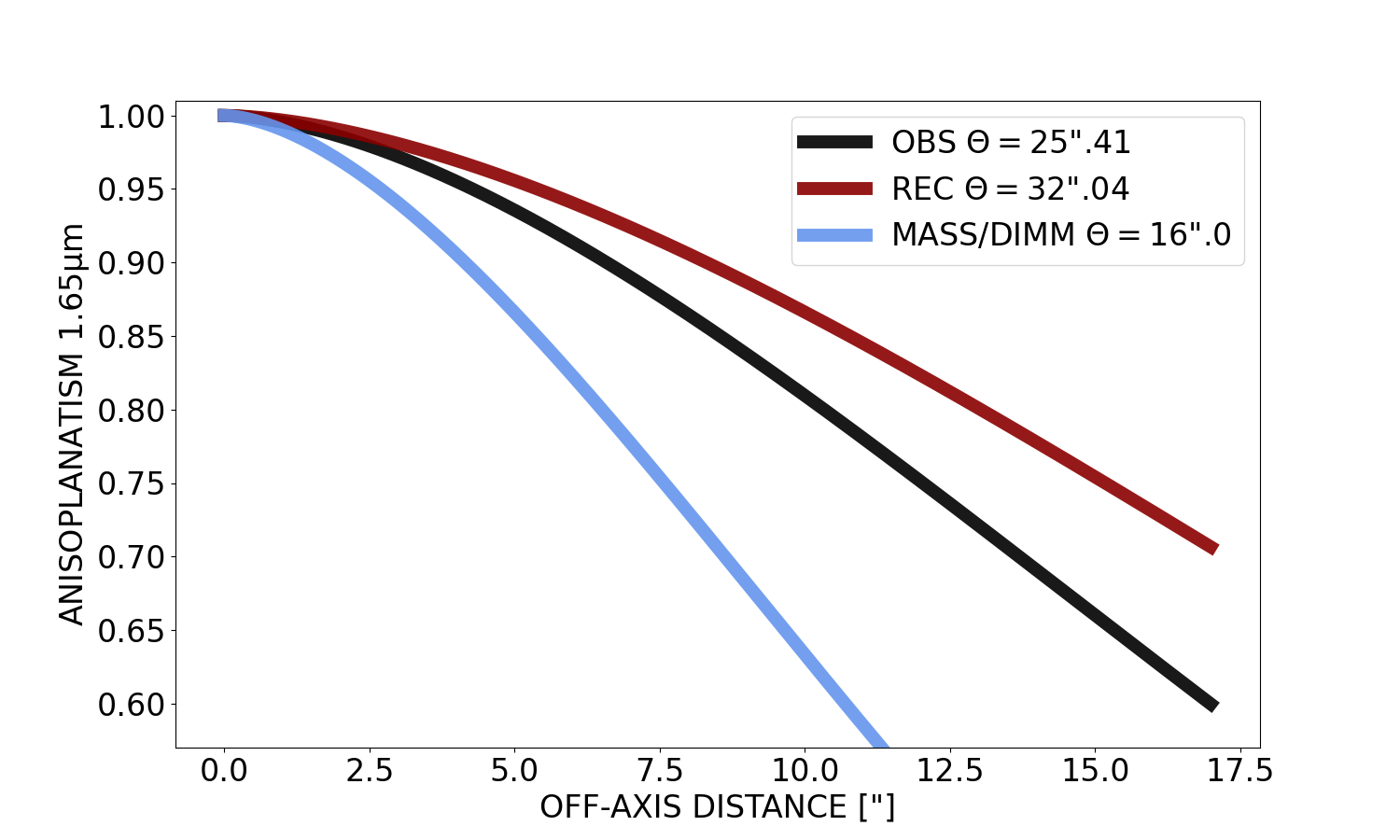}
   \end{tabular}
   \end{center}
   \caption[Anisoplanatism]{ \label{fig:isoa} Anisoplanatism derived from the distribution of measured SR in the ERIS observations (black line) is compared with the anisoplanatism obtained from the reconstructed PSFs (red line). For reference, the same quantity derived from MASS-DIMM measurements, at the same wavelength, is also reported (azure line).}
\end{figure}

As a consistency check, the obtained SR values as a function of off-axis distance have been used to derive an isoplanatic angle\cite{1982JOSA...72...52F,1993ARA&A..31...13B,1999aoa..book.....R}:
\begin{equation*}
    SR(\theta)\propto \exp{\left( \frac{\theta}{\theta_{0}} \right) ^{5/3}}
\end{equation*}
where $\theta_{0}$ is the isoplanatic angle and $\theta$ is the off-axis distance. The isoplanatic angle, in the wavelength of observations, that best reproduces the observed data, results in $\theta_{0,OBS}=25".41$, while using the reconstructed PSFs a $\theta_{0,REC}=32".04$ results. For comparison, MASS-DIMM measurements associated with the observations\footnote{the Fried parameter r$_{0}$ is proportional to $\lambda^{6/5}$\cite{1966JOSA...56.1372F}.} provide a $\theta_{0,1.65{\rm \mu m}}=16".00$. The corresponding anisoplanatisms are reported in Figure~\ref{fig:isoa}. A possible cause for the higher value of $\theta_{0,OBS}$ compared to the MASS-DIMM measurements could be the VLT telescope vibrations: these are fully isoplanatic and intrinsic of the VLT telescope and not to the DIMM which is on a separated mounting on the Paranal platform.
%FIG 3
%%%%%%

The performance of our PSF-R method, in the SCAO off-axis case is reported in Table~\ref{tab:res}. For each SCAO off-axis PSF considered, and the associated reconstructed one, the values of SR, FWHM, and encircled energy of the core are provided. A precision of the reconstruction of the order of $10\%$ is reached in each case. 

%TAB 1
\begin{table}[ht]
\caption{Off-axis PSF-R results for real ERIS SCAO data (first 2 columns) and simulated MICADO ones (last 3 columns). For each dataset, values of SR, FWHM, and encircled energy of the core are provided, along with the effective wavelength of the PSF ($\lambda_{\rm eff}$) and the off-axis distance.} 
\label{tab:res}
\begin{center}   
\begin{tabular}{|l|r|r|r|r|r|}
%% |l|l| to left justify each column entry
%% |c|c| to center each column entry
%% use of \rule[]{}{} below opens up each row
\hline
 & \multicolumn{2}{|c|}{REAL} & \multicolumn{3}{|c|}{SIMULATED} \\
 \cline{2-3} \cline{4-6} 
 & ERIS\_7" & ERIS\_16" & MICADO\_bright & MICADO\_int. & MICADO\_faint \\
\hline
SR$_{\rm OBS}\,[\%]$               & 60   & 45   & 12   & 12   & 7    \\
\hline
SR$_{\rm REC}\,[\%]$               & 50   & 51   & 13   & 12   & 8    \\
\hline
FWHM$_{\rm OBS}\,[mas]$            & 54   & 56   & 13   & 13   & 14   \\
\hline
FWHM$_{\rm REC}\,[mas]$            & 53   & 54   & 12   & 13   & 13   \\
\hline
EE$_{\rm core,OBS}$                & 0.35 & 0.24 & 0.07 & 0.07 & 0.04 \\
\hline
EE$_{\rm core,REC}$                & 0.33 & 0.30 & 0.07 & 0.07 & 0.04 \\
\hline
$\lambda_{\rm eff}\,[{\rm \mu m}]$ & 1.65 & 1.65 & 2.2  & 2.2  & 2.2  \\
\hline
off-axis $["]$                     & 7    & 16   & 30   & 30   & 30   \\
\hline
\end{tabular}
\end{center}
\end{table} 
%%%%%%

It can be noted that the reconstruction accuracy is, on average, higher for MICADO PSFs. This is due, in part, to the total control over the input atmospheric conditions: something that with real data cannot be easily achieved. Still, fundamental sources of errors, intrinsic to the temporal tomography approach, are already included; for example, approximating the initial conditions with a model atmosphere of just $3$ layers for the reconstruction.    

\subsection{Scientifc evaluation of MICADO off-axis PSF-R}
An important activity of the MICADO PSF-R team, complementing the development of the MICADO PSF-R tool, is the scientific evaluation of the reconstructed PSFs\cite{2020SPIE11448E..37S,2022SPIE12185E..41G}. Specifically, we focus on assessing how the PSF-R performances affect the measurement of scientific quantities in selected science cases. 
Our telemetry-only PSF-R approach is particularly convenient in the case of typical extra-galactic observations, where too few (or none) suitable point sources are present in the field of view to characterize the PSF directly from scientific frames\cite{2020SPIE11448E..37S,2022SPIE12185E..0DS,2022JATIS...8c8003S}. For this reason, we use the MICADO off-axis PSF associated with the intermediate AO reference brightness regime, as the template PSF for generating MICADO SCAO simulated observations of a typical z=2 galaxy hosting a population of star-forming regions. Specifically, SimCADO\cite{2016SPIE.9911E..24L} has been used to create MICADO observations, using GALFIT\cite{2010AJ....139.2097P} for both generating the input galaxy model and, later, to perform the morphological and photometric analysis on the obtained images. 

The galaxy model used to generate the simulated MICADO observations samples a large interval of star-forming region sizes (and consequently brightness; see e.g. Ref.~\citenum{2022MNRAS.516.3532M}), but we are primarily interested in exploring the effect of PSF-R in the high spatial-resolution regime, where accurately reconstructing the shape of the PSF, and in particular matching its FWHM, is fundamental. Specifically, the resolution limit of MICADO, in terms of the radius of compact sources at redshift $2$ is about $45$pc, which corresponds to $\sim5.2$mas ($40\%$ of PSF FWHM with a pixel-scale of $4$mas/px). 
%FIG 4
\begin{figure} [tb]
   \begin{center}
   \begin{tabular}{c} %% tabular useful for creating an array of images 
   \includegraphics[width=.7\textwidth]{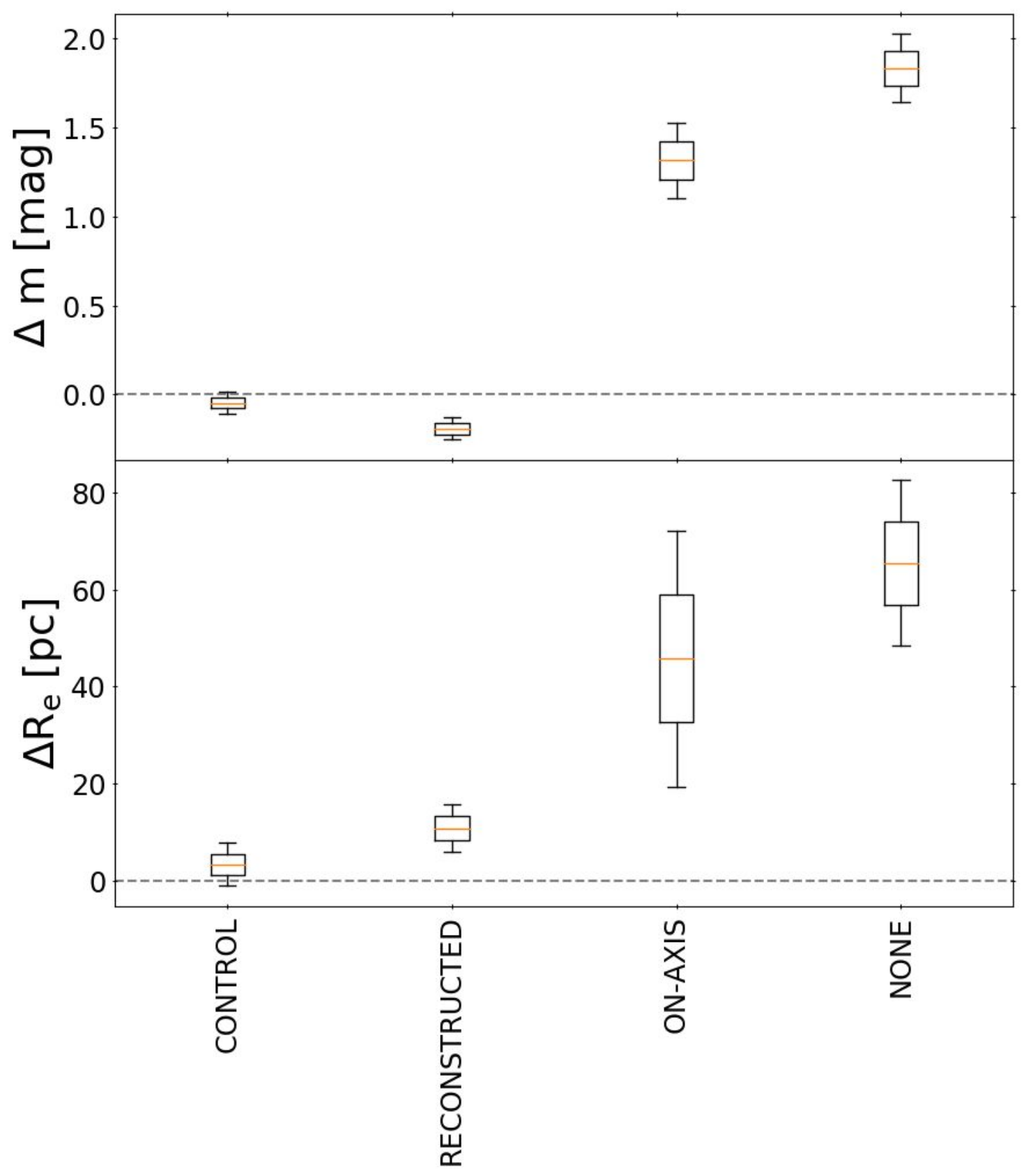}
   \end{tabular}
   \end{center}
   \caption[scieval]{ \label{fig:scieval} Scientific evaluation results: characterization of the size (bottom panel) and luminosity (top panel) of a compact star-forming region in a typical z$=2$ galaxy. The input size of the star-forming region is $45$ pc; the MICADO 30" off-axis PSF is used as a template and the effect of using different PSFs is evaluated. The star-forming region cannot be resolved if no information on the PSF is available (NONE). Moreover, also the associated on-axis MICADO PSF is not a good representation of the actual PSF. In this case, the obtained reconstructed PSF (off-axis) allows the successful recovery of both the size and the luminosity of the star-forming region.}
\end{figure} 
%%%%%%
The results of the analysis are presented in Figure~\ref{fig:scieval}. They indicate that the precision reached by our off-axis PSF-R on MICADO PSFs allows the successful recovery of both the morphological (with an error around $20\%$ in size) and photometric properties of the selected star-forming region.

\section{CONCLUSION}\label{sec:conclusion}
In the context of the MICADO PSF-R, a tomographic method has been recently implemented in the algorithm to allow the reconstruction of off-axis PSFs in SCAO observations, without using any information coming from the scientific images\cite{2023JOSAA..40.1382W}. We used dedicated ERIS@VLT observations (taken in the context of an ESO TTR) of a bright asterism of $3$ stars, to validate the method with real SCAO data.

A precision of the order of $10\%$ can be achieved in SR, FWHM, and encircled energy of the core can be achieved for off-axis distances up to half isoplanatic angle.
The current results are already paving the way to extend the tool for MCAO observations; still, an extended dataset is needed to properly assess the consistency and applicability of the MICADO PSF-R method. For this purpose, the mentioned ESO TTR is aimed at collecting ERIS SCAO data as diverse as possible in terms of atmospheric and AO conditions. All the planned observations will be available to the community.

Finally, our off-axis PSF-R method has been applied also to simulated MICADO PSFs. A 30" off-axis MICADO SCAO PSF has been used to assess the impact of the PSF-R on the characterization of a compact star-forming region in an idealized z=2 galaxy. Encouraging results have been obtained using the reconstructed PSF, both in terms of size of the star-forming region  ($\sim20\%$ error in effective radius) and of its luminosity (of the order of $0.2$mag in integrated magnitude), suggesting that an acceptable precision could be reached with our PSF-R tool. 

\acknowledgments % equivalent to \section*{ACKNOWLEDGMENTS}       
We thank  E. Valenti, M. LeLouarn, J. Vernet, and M. Cirasuolo of ESO for their effort in preparing and coordinating the ERIS technical time requests. MS and AG acknowledge partial support from Bando Ricerca Fondamentale INAF 2023 (Ob. Fu. 1.05.23.04.05, P.I. Simioni).

% References
\bibliography{micado_psfr_offaxis_spie24} % bibliography data in report.bib
\bibliographystyle{spiebib} % makes bibtex use spiebib.bst

\end{document}